\documentclass[review, 1p]{elsarticle}

\usepackage{amsmath,amssymb,bm,booktabs,color}
\usepackage{multirow}
\usepackage{lineno}
\biboptions{sort&compress}

\journal{Physics Letters B}

\begin{document}

\begin{frontmatter}

\title{Nuclear chiral rotation within Relativistic Configuration-interaction Density functional theory}

\author[PKU]{Y. K. Wang}

\author[PKU]{P. W. Zhao\corref{mycorrespondingauthor}}
\ead{pwzhao@pku.edu.cn}

\author[PKU,CIAE]{J. Meng\corref{mycorrespondingauthor}}
\cortext[mycorrespondingauthor]{Corresponding author}
\ead{mengj@pku.edu.cn}

\address[PKU]{State Key Laboratory of Nuclear Physics and Technology, School of Physics, Peking University, Beijing 100871, China}
\address[CIAE]{China Institute of Atomic Energy, Beijing 102413, China}

\begin{abstract}
  The \textit{Re}lativistic \textit{C}onfiguration-interaction \textit{D}ensity functional (ReCD) theory that combines the advantages of large-scale configuration-interaction shell model and relativistic density functional theory is extended to study nuclear chiral rotation.
  The energy spectra and transition probabilities of the chiral doublet bands are reproduced satisfactorily without any free parameters.
  By analyzing the probability amplitudes of the wavefunctions, the significant roles of configuration mixing and four quasiparticle states to the chiral doublets are revealed.
  The evolution from chiral vibration to static chirality are clearly illustrated by the \textit{K plot} and \textit{azimuthal plot}.
  The present investigation provides both microscopic and quantal descriptions for nuclear chirality for the first time and demonstrates the robustness of chiral geometry against the configuration mixing as well as the four quasiparticle states.
\end{abstract}

\begin{keyword}
Chiral rotation \sep rotational symmetry restoration \sep relativistic density functional theory \sep configuration interaction
\end{keyword}

\end{frontmatter}


\section{\label{sec1} Introduction}
Chirality has emerged as a prominent research area in many fields, such as chemistry, biology, and physics.
The chirality in atomic nuclei has garnered great attentions since its first prediction by Frauendorf and Meng in 1997~\cite{Frauendorf1997NuclearPhysicsA131147}.
The predicted topology, namely the mutually perpendicular angular momenta of the valence protons, valence neutrons, and the core, forms left- and right-handed configurations and leads to the spontaneous chiral symmetry breaking in the intrinsic frame.
The restoration of the broken chiral symmetry in the laboratory frame is manifested by the observation of the chiral doublet bands, which consist of a pair of nearly degenerate $\Delta I = 1$ rotational bands with the same parity.
In 2006, a new phenomenon, multiple chiral doublets (M$\chi$D), i.e., more than one pair of chiral doublets in one single nucleus is predicted~\cite{Meng2006Phys.Rev.C037303}.
The evidence of M$\chi$D is confirmed experimentally in 2013~\cite{Ayangeakaa2013Phys.Rev.Lett.172504}. 
The M$\chi$D phenomenon demonstrates the coexistence of nuclear triaxiality and the multiple facets of nuclear chiral rotation.
So far, over 60 chiral doublet bands, including 8 M$\chi$D, have been experimentally reported; see reviews~\cite{Frauendorf2001Rev.Mod.Phys.463514,Meng2010JournalofPhysicsGNuclearandParticlePhysics064025,Meng2016PhysicaScripta053008} and also data compilation~\cite{Xiong2019AtomicDataandNuclearDataTables193225} for more details.

Theoretically, the nuclear chirality has been extensively studied using both phenomenological~\cite{Frauendorf1997NuclearPhysicsA131147,Koike2004Phys.Rev.Lett.172502,Peng2003Phys.Rev.C044324,Zhang2007Phys.Rev.C044307,
Qi2009PhysicsLettersB175180,Chen2018PhysicsLettersB744749,Raduta2016JournalofPhysicsGNuclearandParticlePhysics095107,Tonev2006Phys.Rev.Lett.052501} and microscopic methods~\cite{Dimitrov2000Phys.Rev.Lett.57325735,Chen2013Phys.Rev.C024314,Bhat2012PhysicsLettersB250254,Chen2017Phys.Rev.C051303,
Chen2018PhysicsLettersB211216,Wang2019Phys.Rev.C054303,Olbratowski2004Phys.Rev.Lett.052501,Zhao2017PhysicsLettersB15,Zhao2019Phys.Rev.C054319,
Peng2020PhysicsLettersB135795,Wang2023PhysicsLettersB137923}.
Based on the successful relativistic density functional theory (DFT)~\cite{Meng2016}, the three-dimensional tilted axis cranking model (3D-TAC-DFT)~\cite{Zhao2017PhysicsLettersB15} with core polarization and nuclear currents considered self-consistently has received more attentions.
Up to now, the relativistic 3D-TAC-DFT has been extended to study the M$\chi$D~\cite{Zhao2017PhysicsLettersB15}, the nuclear chiral conundrum~\cite{Zhao2019Phys.Rev.C054319}, and the superfluidity effects on nuclear chiral rotation~\cite{Wang2023PhysicsLettersB137923}.
In these studies, however, the angular momenta and transition probabilities are treated in semiclassical ways, and only the lower-energy band of the chiral doublets can be obtained.

To describe the lower- and upper-energy bands of the doublets simultaneously, the nuclear chirality is investigated dynamically within the time-dependent relativistic 3D-TAC-DFT~\cite{Ren2022Phys.Rev.CL011301}.
The energy splitting between the doublet bands can be reproduced and explained by the chiral precession.
However, the time-dependent relativistic 3D-TAC-DFT is still within the framework of mean-field approximation.

Recently, the \textit{Re}lativistic \textit{C}onfiguration-interaction \textit{D}ensity functional (ReCD) theories in axial~\cite{Zhao2016Phys.Rev.C041301} and triaxial~\cite{Wang2023arXivpreprintarXiv2304.12009} cases are developed.
The basic idea of the ReCD theory is the following.
Firstly, the relativistic DFT calculation is performed to obtain a self-consistent solution, which corresponds to the minimum of the potential energy surface and includes already important physics.
Secondly, the configuration space including various quasiparticle excitation states is constructed based on the self-consistent solution.
Thirdly, the broken rotational symmetry for states in the configuration space is restored by the angular momentum projection, and a set of projected bases with good angular momentum is obtained.
Finally, a shell-model calculation, namely diagonalizing the Hamiltonian in symmetry restored  space expanded by the projected bases,  is carried out to consider the correlations required to describe nuclear spectroscopic properties.
It is clear that the ReCD theory combines the advantages of large-scale configuration-interaction shell model and relativistic DFT and, thus, provides a promising tool to study the properties of nuclear chirality in both microscopic and quantal ways.

In this work, we report the first application of the ReCD theory for the nuclear chirality.
The chiral doublet bands in the odd-odd nucleus $^{130}$Cs~\cite{Starosta2001Phys.Rev.Lett.971974} are investigated as an example.

\section{\label{sec2} Theoretical Framework}
In the ReCD theory, the nuclear wavefunction formulated in the laboratory frame is expressed as~\cite{Zhao2016Phys.Rev.C041301,Wang2022Phys.Rev.C054311,Wang2023arXivpreprintarXiv2304.12009}
\begin{equation}\label{eq1}
  |\Psi^{IM}_\sigma\rangle = \sum_{K\kappa}F^{I\sigma}_{K\kappa}\hat{P}^I_{MK}|\Phi_\kappa\rangle,
\end{equation}
where $\hat{P}^I_{MK}$ is the three-dimensional angular momentum operator~\cite{Ring2004}, $\hat{P}^I_{MK}|\Phi_\kappa\rangle$ is the projected basis with good angular momentum $I$ and $M$, and $|\Phi_\kappa\rangle$ represents a certain intrinsic state in the configuration space, which up to four quasiparticle states for odd-odd nuclei is constructed as~\cite{Wang2023arXivpreprintarXiv2304.12009}
\begin{equation}\label{eq2}
  \{\hat{\beta}^\dag_{\pi_0}\hat{\beta}^\dag_{\nu_0}|\Phi_0\rangle,\hat{\beta}^\dag_{\pi_i}\hat{\beta}^\dag_{\nu_j}|\Phi_0\rangle,
  \hat{\beta}^\dag_{\pi_i}\hat{\beta}^\dag_{\nu_j}\hat{\beta}^\dag_{\pi_k}\hat{\beta}^\dag_{\pi_l}|\Phi_0\rangle,
  \hat{\beta}^\dag_{\pi_i}\hat{\beta}^\dag_{\nu_j}\hat{\beta}^\dag_{\nu_k}\hat{\beta}^\dag_{\nu_l}|\Phi_0\rangle\}.
\end{equation}
Here, $\hat{\beta}^\dag_\pi$ and $\hat{\beta}_\nu^\dag$ are the quasiparticle (qp) creation operators for protons and neutrons, respectively.
Among all the states in Eq.~\eqref{eq2}, $|\Phi_{\pi_0\nu_0}\rangle\equiv\hat{\beta}^\dag_{\pi_0}\hat{\beta}^\dag_{\nu_0}|\Phi_0\rangle$ has the lowest intrinsic total energy and it is obtained by iteratively solving the triaxial relativistic Hartree-Bogoliubov (TRHB) equation~\cite{Meng2016}.
To ensure the correct number parity for the odd-odd nucleus, the proton ($\pi_0$) and the neutron ($\nu_0$) orbits with the lowest qp energies are blocked during the iterative calculations~\cite{Ring2004}.
The coefficients $F^{I\sigma}_{K\kappa}$ in Eq.~\eqref{eq1} play the role of variational parameters and are determined by the following generated eigenvalue equation
\begin{equation}\label{eq3}
  \sum_{K\kappa}\{H^I_{K'\kappa'K\kappa} - E^{I\sigma}N^I_{K'\kappa'K\kappa}\}F^{I\sigma}_{K\kappa} = 0.
\end{equation}
The Hamiltonian matrix element and the norm matrix element are defined as
\begin{equation}\label{eq4}
  H^I_{K'\kappa'K\kappa} = \langle\Phi_{\kappa'}|\hat{H}\hat{P}^I_{K'K}|\Phi_\kappa\rangle,\quad
  N^I_{K'\kappa'K\kappa} = \langle\Phi_{\kappa'}|\hat{P}^I_{K'K}|\Phi_\kappa\rangle,
\end{equation}
and are evaluated by the Pfaffian algorithms proposed in Refs.~\cite{Hu2014PhysicsLettersB162166,Carlsson2021Phys.Rev.Lett.172501}.
The Hamiltonian $\hat{H}$ is derived from a universal relativistic Lagrangian density by the Legendre transformation~\cite{Zhao2016Phys.Rev.C041301}.
Once $F^{I\sigma}_{K\kappa}$ are obtained, one can calculate the physical observables associated with the nuclear chirality, including the $E2$ and $M1$ transition probabilities.

It is known that the projected bases \{$\hat{P}^I_{MK}|\Phi_\kappa\rangle$\} are not orthogonal and the coefficients $F^{I\sigma}_{K\kappa}$ should not be interpreted as the probability amplitudes for the state $|\Phi_\kappa\rangle$.
To obtain the probability amplitudes, one needs to construct the following collective wavefunctions~\cite{Zhao2016Phys.Rev.C041301}
\begin{equation}\label{eq5}
  g^{I\sigma}_{K\kappa} = \sum_{K'\kappa'} (N^I)^{1/2}_{K\kappa K'\kappa'}F^{I\sigma}_{K'\kappa'}
\end{equation}
with $(N^I)^{1/2}_{K\kappa K'\kappa'}$ the matrix element of the square root of the norm matrix in Eq.~\eqref{eq4}.
The probability amplitude for each state $|\Phi_\kappa\rangle$ in the configuration space is then expressed as
\begin{equation}\label{eq6}
  G_\kappa^{I\sigma} = \sum_K |g^{I\sigma}_{K\kappa}|^2.
\end{equation}
The $G_\kappa^{I\sigma}$ will be used to analyze the dominant configurations of the wavefunction $|\Psi^{IM}_\sigma\rangle$ and examine the structural evolution of chiral doublet bands with the total angular momentum $I$.
The collective wavefunctions in Eq.~\eqref{eq5} can also be used in the calculations of the \textit{K plot} and the \textit{azimuthal plot}, which have been introduced in Ref.~\cite{Chen2017Phys.Rev.C051303} to illustrate the chiral geometry of the chiral doublet bands.
The \textit{K plot} represents the probability distributions of the components of the total angular momentum on the three intrinsic axes~\cite{Chen2017Phys.Rev.C051303},
\begin{equation}\label{eq7}
  P^{I\sigma}(|K|) = \sum_\kappa|g^{I\sigma}_{K\kappa}|^2 + |g^{I\sigma}_{-K\kappa}|^2.
\end{equation}
The \textit{azimuthal plot} represents the probability distributions of the orientation of the total angular momentum on the intrinsic $(\theta,\phi)$ plane~\cite{Chen2017Phys.Rev.C051303},
\begin{equation}\label{eq8}
  \mathcal{P}^{I\sigma}(\theta,\phi) = \sum_\kappa \int_0^{2\pi} d\psi'|W^{I\sigma}_\kappa(\psi',\theta,\pi-\phi)|^2,
\end{equation}
where the integrand $W^{I\sigma}_\kappa(\psi',\theta,\pi-\phi)$ reads
\begin{equation}\label{eq9}
  W^{I\sigma}_\kappa(\psi',\theta,\pi-\phi) = \sqrt{\frac{2I+1}{8\pi^2}}\sum_K g^{I\sigma}_{K\kappa} D^{I\ast}_{IK}(\psi',\theta,\pi-\phi).
\end{equation}
Here, $\theta$ is the angle between the total angular momentum and the long ($l$) axis, and $\phi$ is the angle between the projection of the total angular momentum on the intermediate-short ($i$-$s$) plane and $i$ axis.

\section{\label{sec3} Results and discussion}
In the present work, the chiral doublet bands (denoted as Band A and Band B) in $^{130}$Cs are studied using the ReCD theory.
The point-coupling Lagrangian PC-PK1~\cite{Zhao2010Phys.Rev.C054319} is adopted to derive the Hamiltonian $\hat{H}$ in Eq.~\eqref{eq4} and the TRHB equation.
A finite-range separable pairing force with strength $G = 728$ MeV fm$^3$~\cite{Tian2009PhysicsLettersB4450} is used to treat the pairing correlations.
The TRHB equation is solved in the three-dimensional harmonic oscillator basis~\cite{Niksic2014ComputerPhysicsCommunications18081821} with 10 major shells.
By solving the TRHB equation iteratively, it is found that the deformation parameters $(\beta,\gamma)$ for $^{130}$Cs are $(0.20,21^\circ)$.
Similar to Refs.~\cite{Zhao2016Phys.Rev.C041301,Wang2022Phys.Rev.C054311,Wang2023arXivpreprintarXiv2304.12009}, the dimension of the configuration space is truncated with a qp excitation energy cutoff $E_{\mathrm{cut}}$.
The $E_{\mathrm{cut}}$ = 5.0 MeV is adopted in the present calculation.
The resultant configuration space is sufficient to obtain convergent results for the spectroscopic properties of $^{130}$Cs.
The calculations are free of additional parameters.

\begin{figure}[htbp]
  \centering
  \includegraphics[width=0.5\textwidth]{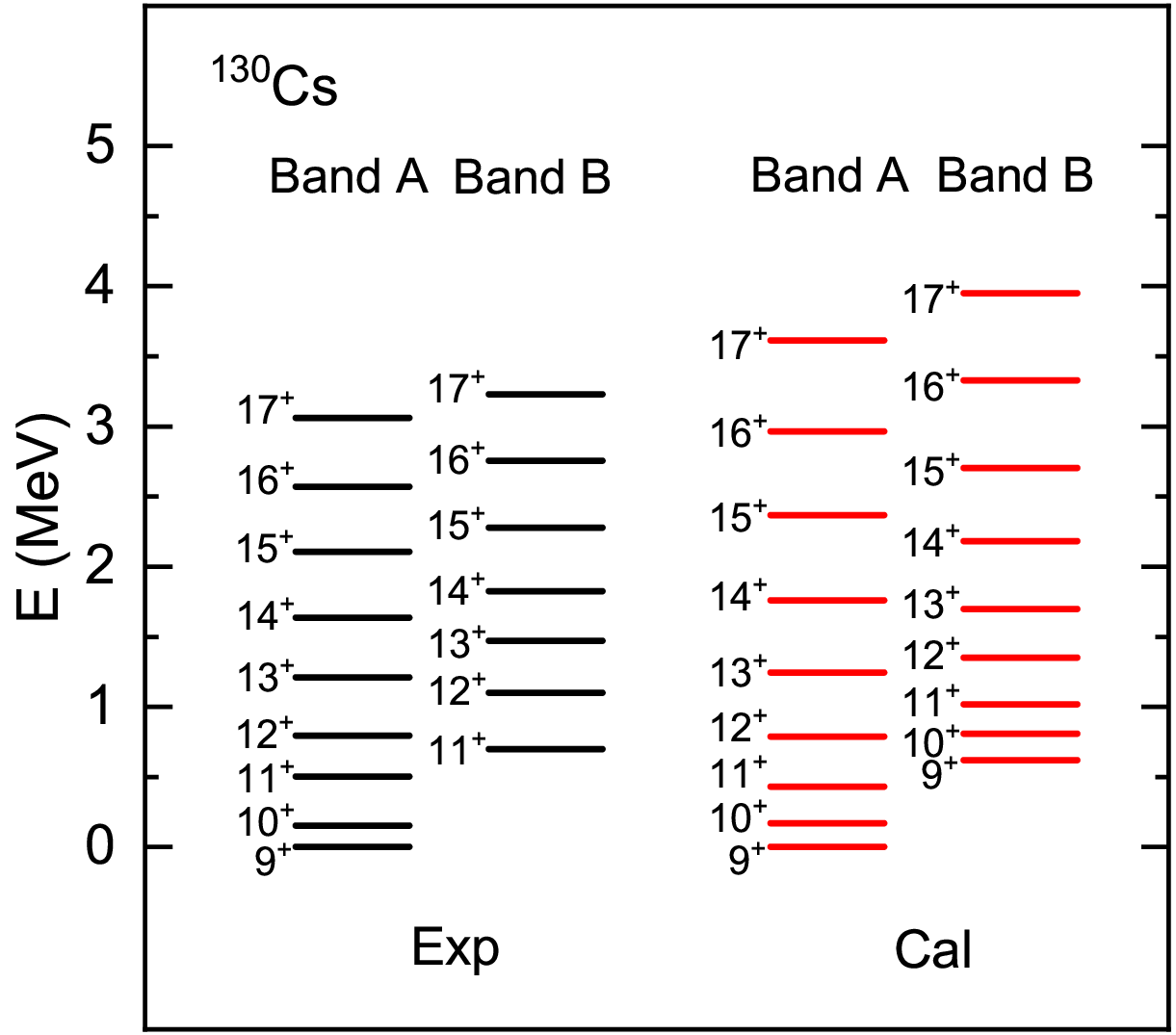}
  \caption{(Color online) Energy spectra of the chiral doublet bands in $^{130}$Cs calculated by the ReCD theory (right panel) in comparison with data~\cite{Starosta2001Phys.Rev.Lett.971974} (left panel).
  The energy levels are shifted by taking the level $9^+$ in Band A as a reference.}
  \label{Spectra}
\end{figure}
The calculated energy spectra for Bands A and B in $^{130}$Cs and their comparison with the data~\cite{Starosta2001Phys.Rev.Lett.971974} are shown in Fig.~\ref{Spectra}.
The predicted spectra, including the near degeneracy between Bands A and B, agree satisfactorily with the data.
In more detail, the energy levels with $I\geq 15\hbar$ are slightly overestimated.
Such overestimation might be alleviated by considering states beyond the four-qp configurations.

\begin{figure}[htbp]
  \centering
  \includegraphics[width=0.5\textwidth]{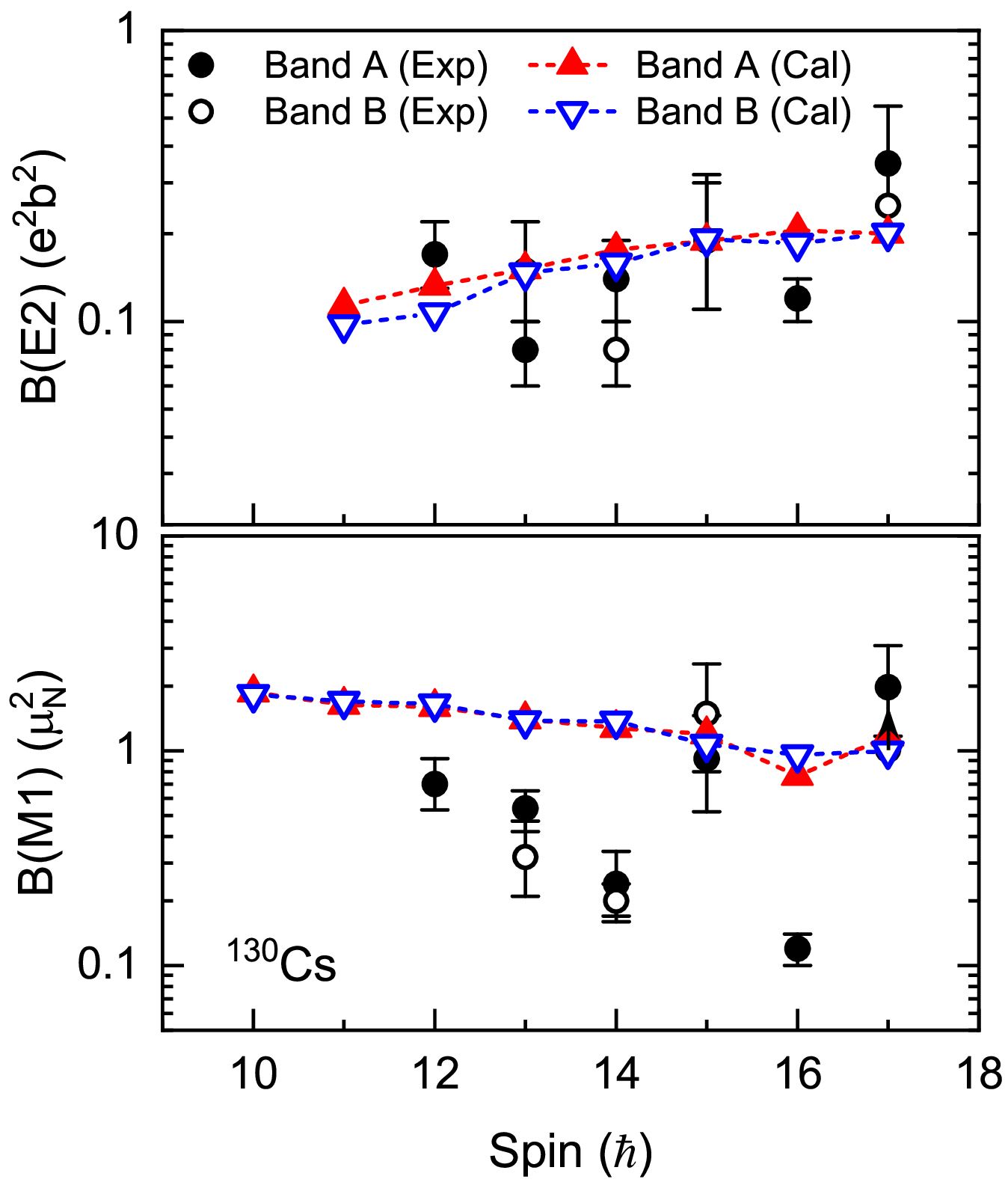}
  \caption{(Color online) $E2$ and $M1$ transition probabilities of Band A and Band B calculated by the ReCD theory, in comparison with the experimental data~\cite{LieLin2009ChinesePhysicsC173}.}
  \label{Transition}
\end{figure}
The theoretical $E2$ and $M1$ transition probabilities for the chiral doublets in $^{130}$Cs are depicted in Fig.~\ref{Transition}, in comparison with available data~\cite{LieLin2009ChinesePhysicsC173}.
The predicted $B(E2)$ values reproduce well the experimental data, and they are similar for Bands A and B, as expected for the chiral doublets~\cite{Koike2004Phys.Rev.Lett.172502,Zhang2007Phys.Rev.C044307}.
Note that there is no need to introduce the effective charge when calculating the $E2$ transition probabilities in the ReCD theory, as demonstrated in Refs.~\cite{Zhao2016Phys.Rev.C041301,Wang2022Phys.Rev.C054311}.
The $B(M1)$ values are somewhat overestimated for states with $I\leq 14\hbar$.
The predicted staggering phenomenon of $B(M1)$ values for Band A at $I = 16\hbar$ is also weaker than the data.
It is known that the staggering of $B(M1)$ values for chiral doublets depends sensitively on the triaxial deformation $\gamma$, and its amplitude decreases significantly as $\gamma$ deviates from $30^\circ$~\cite{Qi2009Phys.Rev.C041302}.
The weaker staggering of $B(M1)$ values predicted here may indicate that the $\gamma$ value of $^{130}$Cs is slightly underestimated by the relativistic DFT.

\begin{figure}[htbp]
  \centering
  \includegraphics[width=0.5\textwidth]{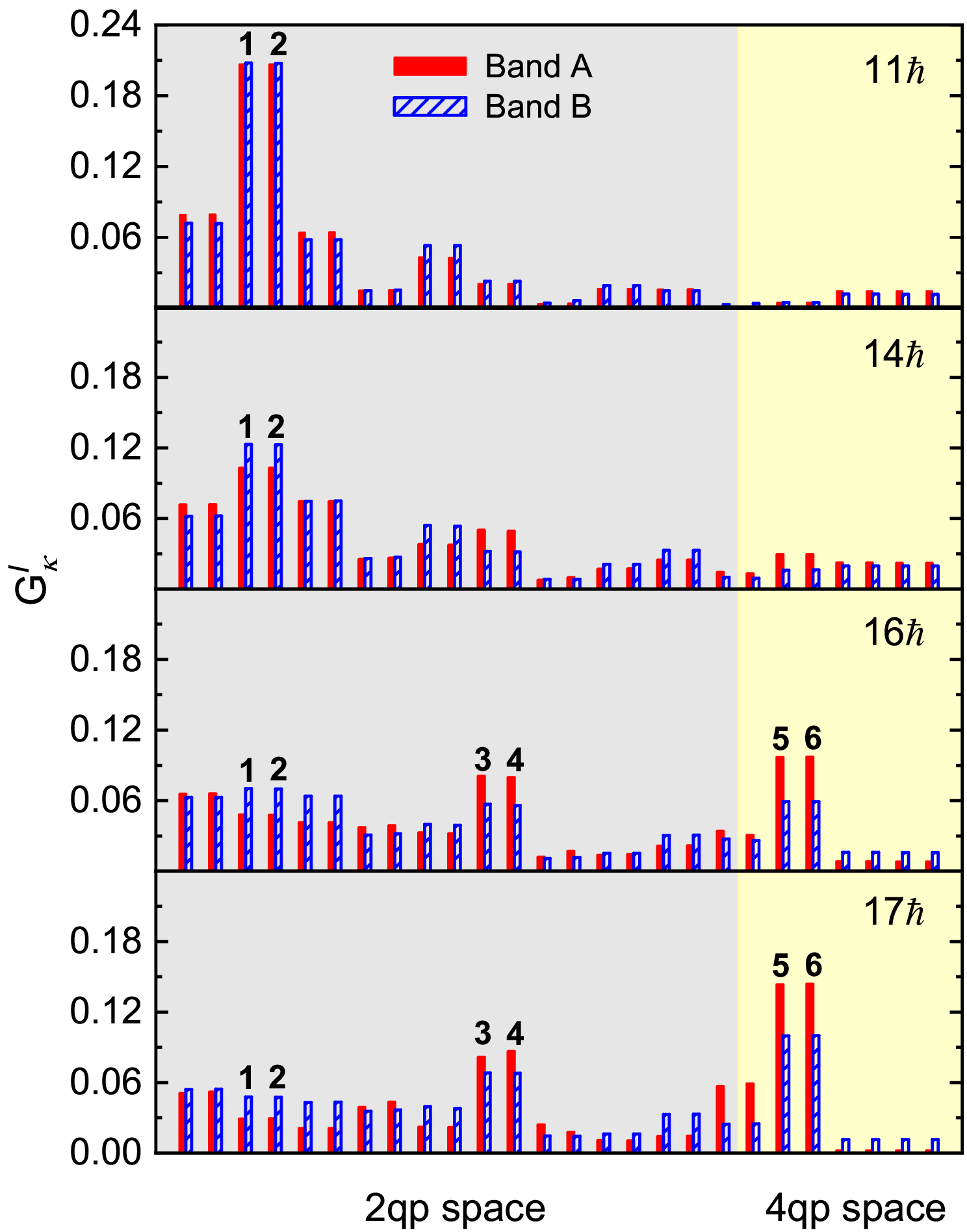}
  \caption{(Color online) The probability amplitudes $G^I_\kappa$ for configurations that play important roles for describing Bands A and B.
  Here, configurations whose contributions to the doublet bands larger than 1\% are listed.
  The most dominant configurations are marked as ``1", ``2", ``3", ``4", ``5", and ``6", and their detailed information are shown in Table~\ref{Configuration}.}
  \label{Wavefunction}
\end{figure}

\begin{table}[htbp]
  \centering
  \caption{Detailed information of the most dominant configurations for Bands A and B.
  Here, $\Omega$ represents the projection of the spin of the $h_{11/2}$ qp orbital on the quantization axis, while $\pi$ and $\nu$ indicate protons and neutrons.}
  \begin{tabular}{cc}
  \toprule
  Labels & Configurations \\
  \midrule
  $1$ & $\pi(h_{11/2;\Omega=1/2})\otimes\nu(h_{11/2;\Omega=9/2})$\\
  $2$ & $\pi(h_{11/2;\Omega=1/2})\otimes\nu(h_{11/2;\Omega=-9/2})$\\
  $3$ & $\pi(h_{11/2;\Omega=1/2})\otimes\nu(h_{11/2;\Omega=5/2})$\\
  $4$ & $\pi(h_{11/2;\Omega=1/2})\otimes\nu(h_{11/2;\Omega=-5/2})$\\
  $5$ & $\pi(h_{11/2;\Omega=1/2})\otimes\nu(h_{11/2;\Omega=9/2}h_{11/2;\Omega=7/2}h_{11/2;\Omega=-7/2})$\\
  $6$ & $\pi(h_{11/2;\Omega=1/2})\otimes\nu(h_{11/2;\Omega=-9/2}h_{11/2;\Omega=7/2}h_{11/2;\Omega=-7/2})$\\
  \bottomrule
  \end{tabular}\label{Configuration}
\end{table}
To pin down the dominating configurations and examine the structural evolution of the chiral doublets in $^{130}$Cs, the probability amplitudes $G_\kappa^{I}$ obtained from Eq.~\eqref{eq6} are shown in Fig.~\ref{Wavefunction}.
Here, we list only the configurations whose contributions to Bands A and B are larger than 1\%.
The most dominant configurations are marked as ``1", ``2", ``3", ``4", ``5", and ``6", and their detailed information are shown in Table~\ref{Configuration}.
The dominant configurations for Bands A and B are similar, as expected for nuclear chiral rotation.
For $I < 16\hbar$, the 2qp states $\pi(h_{11/2;\Omega=1/2})\otimes\nu(h_{11/2;\Omega=9/2})$ and $\pi(h_{11/2;\Omega=1/2})\otimes\nu(h_{11/2;\Omega=-9/2})$ are dominant configurations.
For $I \geq 16\hbar$, the 2qp states $\pi(h_{11/2;\Omega=1/2})\otimes\nu(h_{11/2;\Omega=5/2})$ and $\pi(h_{11/2;\Omega=1/2})\otimes\nu(h_{11/2;\Omega=-5/2})$, and the 4qp states $\pi(h_{11/2;\Omega=1/2})\otimes\nu(h_{11/2;\Omega=9/2}h_{11/2;\Omega=7/2}h_{11/2;\Omega=-7/2})$ and $\pi(h_{11/2;\Omega=1/2})\otimes\nu(h_{11/2;\Omega=-9/2}h_{11/2;\Omega=7/2}h_{11/2;\Omega=-7/2})$ take over, indicating the important roles of qp configuration mixing and 4qp states for describing the chiral doublets in $^{130}$Cs.

\begin{figure}[htbp]
  \centering
  \includegraphics[width=0.7\textwidth]{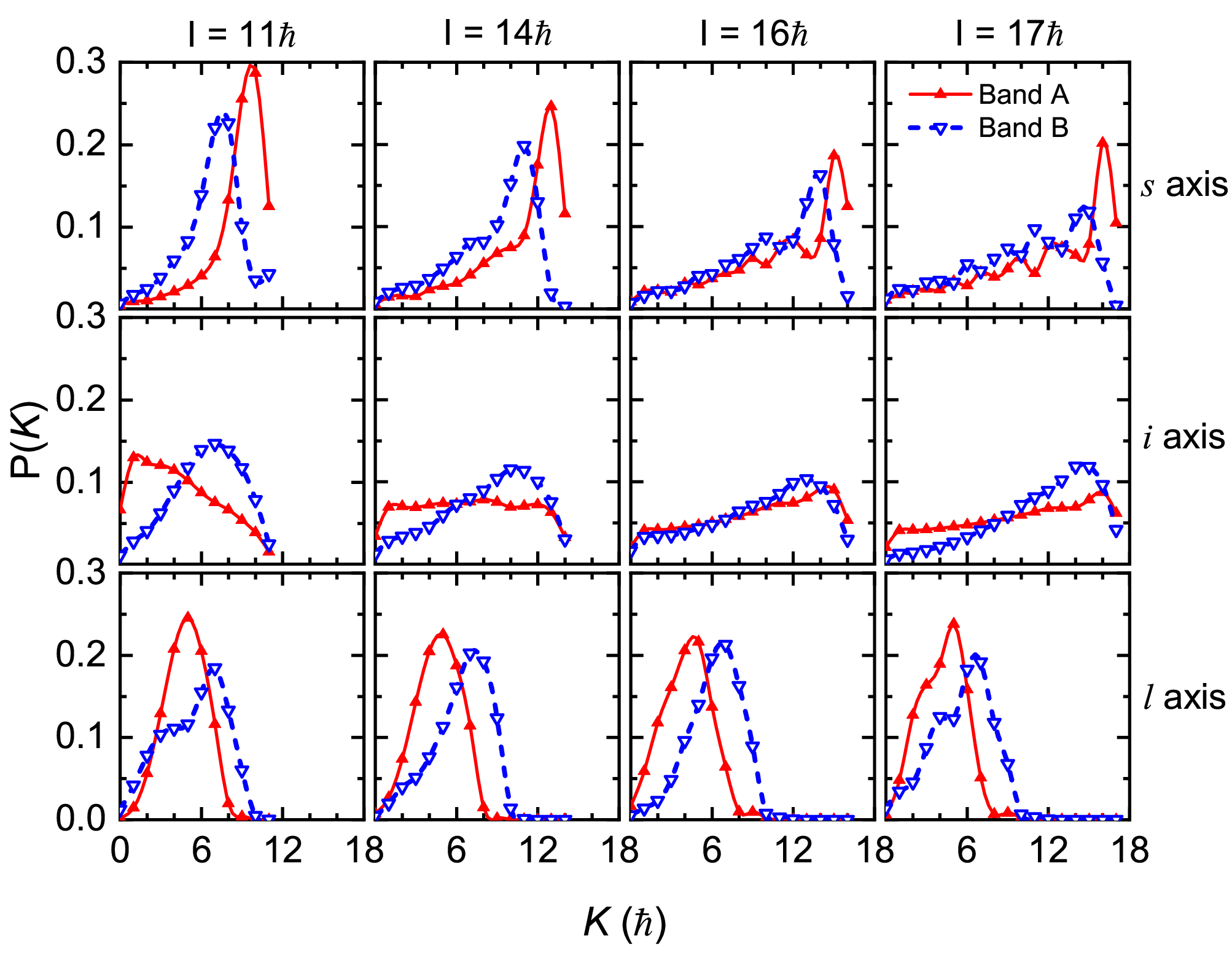}
  \caption{(Color online) The \textit{K plot}, i.e., the $K$ distributions of the angular momentum on the three principal axes, for Band A and Band B at $I = 11\hbar$, $14\hbar$, $16\hbar$, and $17\hbar$.
  The $K$ distributions on \textit{short}, \textit{intermediate}, and \textit{long} axes are shown respectively in the first, second, and third rows.}
  \label{Kplot}
\end{figure}
The chiral geometry for chiral doublets can be illustrated by the \textit{K plot} and the \textit{azimuthal plot}, as proposed in Ref.~\cite{Chen2017Phys.Rev.C051303}.
The \textit{K plot}, i.e., the probability distributions of the components of the total angular momentum on the three intrinsic axes, for Bands A and B at $I = 11\hbar$, $14\hbar$, $16\hbar$, and $17\hbar$ are shown in Fig.~\ref{Kplot}.
The evolution from the chiral vibration to the static chirality can be clearly seen from the changes of \textit{K} distributions on the \textit{i} axis.
At $I = 11\hbar$, the \textit{K} distribution for Band A exhibits a peak at $K \approx 0\hbar$, and the one for Band B peaks at $K\approx 8\hbar$.
This is the typical feature of zero- and one-phonon states and can be interpreted as chiral vibration with respect to the \textit{l}-\textit{s} plane, as demonstrated in Refs.~\cite{Chen2017Phys.Rev.C051303,Chen2018PhysicsLettersB211216,Wang2019Phys.Rev.C054303}.
At $I = 14\hbar$, the \textit{K} distribution becomes rather flat for Band A.
This indicates that the total angular momentum of Band A begins to deviate from the \textit{l}-\textit{s} plane and the collective rotation around \textit{i} axis develops.
At $I = 16\hbar$ and $17\hbar$, the \textit{K} distributions for both Bands A and B peak at $K\approx16\hbar$.
The similar \textit{K} distributions for Bands A and B suggest the appearance of static chirality.

\begin{figure}[htbp]
  \centering
  \includegraphics[width=0.7\textwidth]{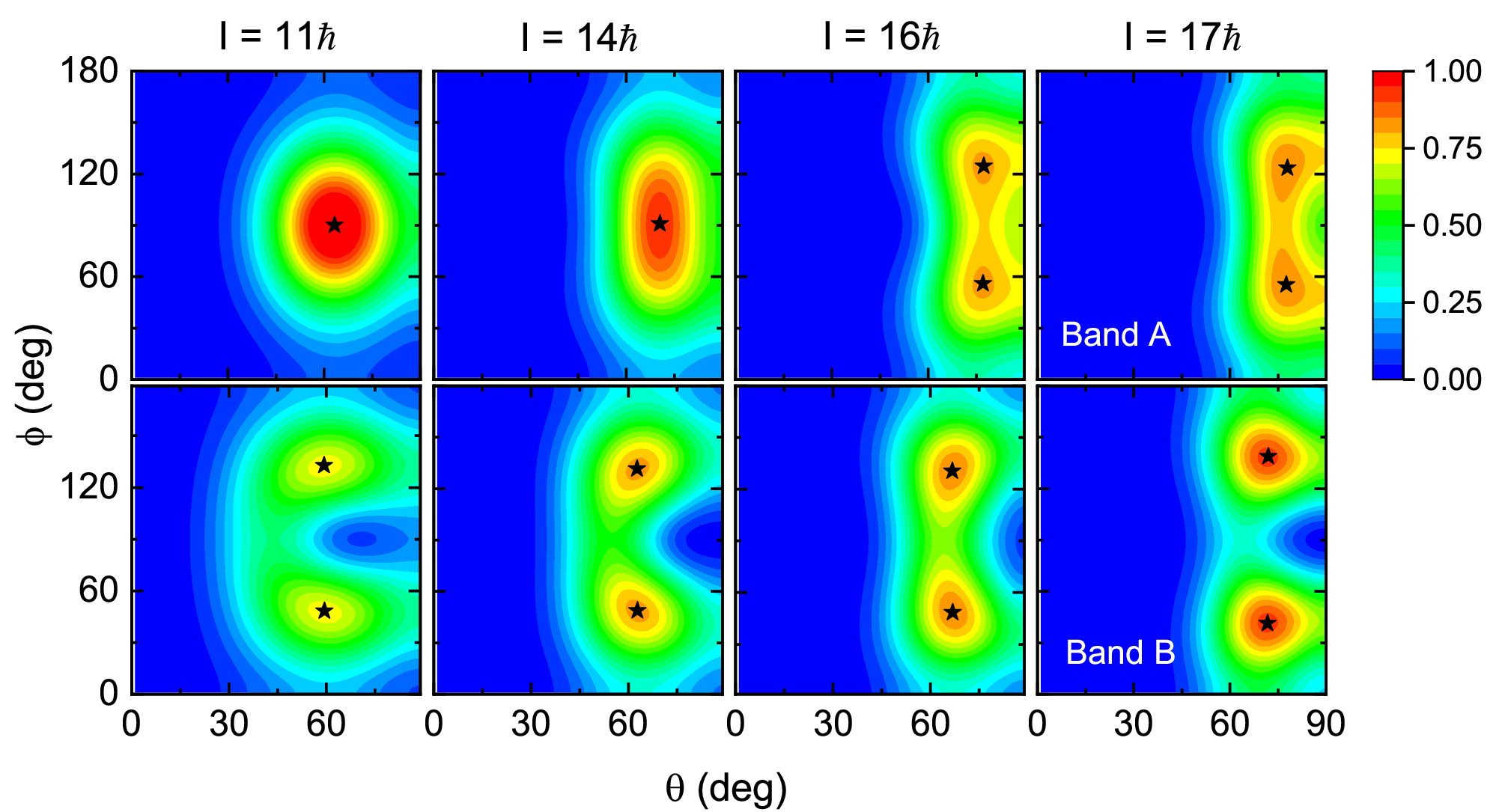}
  \caption{(Color online) The \textit{azimuthal plot}, i.e., the probability distribution profiles for the orientation of the angular momentum on the intrinsic $(\theta,\phi)$ plane for Band A (upper panel) and Band B (lower panel) at $I = 11\hbar$, $14\hbar$, $16\hbar$, and $17\hbar$.
  The black stars represent the positions of local minima.}
  \label{Aplot}
\end{figure}
The chiral geometry can also be examined by the \textit{azimuthal plot}.
The \textit{azimuthal plot}, i.e., the probability distribution profiles for the orientation of the angular momentum on the intrinsic $(\theta,\phi)$ plane, for Bands A and B at $I = 11\hbar$, $14\hbar$, $16\hbar$, and $17\hbar$ are shown in Fig.~\ref{Aplot}.

At $I = 11\hbar$, the \textit{azimuthal plot} for Band A has one single peak at $(\theta,\phi) = (63^\circ,90^\circ)$, which means the total angular momentum of Band A orientates mainly at \textit{l}-\textit{s} plane and corresponds to a planar rotation.
This is in accordance with the expectation for the zero-phonon state~\cite{Chen2017Phys.Rev.C051303,Chen2018PhysicsLettersB211216}.
The \textit{azimuthal plot} for Band B exhibits two peaks at $(\theta,\phi) = (59^\circ,48^\circ)$ and $(\theta,\phi) = (59^\circ,132^\circ)$, together with a node at $(\theta,\phi) = (70^\circ, 90^\circ)$, supporting the interpretation of the one-phonon vibration.
Therefore, the picture of chiral vibration is clearly demonstrated, and this is also consistent with the same probability amplitudes $G^I_\kappa$ for Bands A and B at $I = 11\hbar$, as shown in Fig.~\ref{Wavefunction}.

At $I = 14\hbar$, there remains one single peak in the \textit{azimuthal plot} for Band A, but the corresponding probability distribution profile along $\phi$ direction is very flat.
This means the probability of the total angular momentum deviating from $\phi = 0^\circ$ (\textit{l}-\textit{s} plane) begins to
increase.
Such phenomenon is consistent with the result revealed by the \textit{K} distribution on the \textit{i} axis at $I = 14\hbar$, as shown in Fig.~\ref{Kplot}.

At $I = 16\hbar$ and $17\hbar$, two peaks corresponding to aplanar orientations of the total angular momentum are found, i.e., $(\theta,\phi) \approx (77^\circ, 55^\circ)$ and $(77^\circ, 125^\circ)$ for Band A, and $(\theta,\phi) \approx (69^\circ, 45^\circ)$ and $(69^\circ, 135^\circ)$ for Band B.
These features suggest the realization of static chirality.
Moreover, the non-vanishing distribution at $\theta = 90^\circ$ and $\phi = 90^\circ$ reflects the tunneling between the left- and right-handed configurations, which explains also the slight differences of the probability amplitudes $G^I_\kappa$ between Bands A and B, as shown in Fig.~\ref{Wavefunction}.
The chiral geometry illustrated by the \textit{K plot} and the \textit{azimuthal plot} is thus confirmed to be robust against the configuration mixing and the four qp states.

\section{\label{sec4} Summary}
In summary, the ReCD theory, i.e., the \textit{Re}lativistic \textit{C}onfiguration-interaction \textit{D}ensity functional theory, that combines the advantages of large-scale configuration-interaction shell model and relativistic density functional theory is extended to study the chiral rotation in $^{130}$Cs.
Without any free parameters, the energy spectra and transition properties of the chiral doublets are reproduced satisfactorily.
By calculating the probability amplitudes for states in the configuration space, the composition of the wavefunctions is analyzed.
It is found that the quasiparticle configuration mixing and four quasiparticle states play important roles for the chiral doublets.
The chiral geometry of the doublets is illustrated in terms of the \textit{K plot} and \textit{azimuthal plot}, from which the evolution from chiral vibration to static chirality with the total angular momentum is clearly seen.
The present investigation provides both microscopic and quantal descriptions for nuclear chirality for the first time and demonstrates the robustness of chiral geometry against the configuration mixing and the four quasiparticle states.

\section*{Acknowledgments}
This work was partly supported by the National Natural Science Foundation of China (Grants No. 12141501, No. 12105004, No. 12070131001, No. 11935003, and No. 11975031), the China Postdoctoral Science Foundation under Grant No. 2020M680183, the High-performance Computing Platform of Peking University, and State Key Laboratory of Nuclear Physics and Technology, Peking University.



\begin{thebibliography}{39}
\expandafter\ifx\csname natexlab\endcsname\relax\def\natexlab#1{#1}\fi
\providecommand{\url}[1]{\texttt{#1}}
\providecommand{\href}[2]{#2}
\providecommand{\path}[1]{#1}
\providecommand{\DOIprefix}{doi:}
\providecommand{\ArXivprefix}{arXiv:}
\providecommand{\URLprefix}{URL: }
\providecommand{\Pubmedprefix}{pmid:}
\providecommand{\doi}[1]{\href{http://dx.doi.org/#1}{\path{#1}}}
\providecommand{\Pubmed}[1]{\href{pmid:#1}{\path{#1}}}
\providecommand{\bibinfo}[2]{#2}
\ifx\xfnm\relax \def\xfnm[#1]{\unskip,\space#1}\fi
\bibitem[{Frauendorf and Meng(1997)}]{Frauendorf1997NuclearPhysicsA131147}
\bibinfo{author}{S.~Frauendorf}, \bibinfo{author}{J.~Meng},
  \bibinfo{journal}{Nucl. Phys. A} \bibinfo{volume}{617} (\bibinfo{year}{1997})
  \bibinfo{pages}{131--147}.
\bibitem[{Meng et~al.(2006)Meng, Peng, Zhang, and
  Zhou}]{Meng2006Phys.Rev.C037303}
\bibinfo{author}{J.~Meng}, \bibinfo{author}{J.~Peng}, \bibinfo{author}{S.~Q.
  Zhang}, \bibinfo{author}{S.-G. Zhou}, \bibinfo{journal}{Phys. Rev. C}
  \bibinfo{volume}{73} (\bibinfo{year}{2006}) \bibinfo{pages}{037303}.
\bibitem[{Ayangeakaa et~al.(2013)Ayangeakaa, Garg, Anthony, Frauendorf, Matta,
  Nayak, Patel, Chen, Zhang, Zhao, Qi, Meng, Janssens, Carpenter, Chiara,
  Kondev, Lauritsen, Seweryniak, Zhu, Ghugre, and
  Palit}]{Ayangeakaa2013Phys.Rev.Lett.172504}
\bibinfo{author}{A.~D. Ayangeakaa}, \bibinfo{author}{U.~Garg},
  \bibinfo{author}{M.~D. Anthony}, \bibinfo{author}{S.~Frauendorf},
  \bibinfo{author}{J.~T. Matta}, \bibinfo{author}{B.~K. Nayak},
  \bibinfo{author}{D.~Patel}, \bibinfo{author}{Q.~B. Chen},
  \bibinfo{author}{S.~Q. Zhang}, \bibinfo{author}{P.~W. Zhao},
  \bibinfo{author}{B.~Qi}, \bibinfo{author}{J.~Meng}, \bibinfo{author}{R.~V.~F.
  Janssens}, \bibinfo{author}{M.~P. Carpenter}, \bibinfo{author}{C.~J. Chiara},
  \bibinfo{author}{F.~G. Kondev}, \bibinfo{author}{T.~Lauritsen},
  \bibinfo{author}{D.~Seweryniak}, \bibinfo{author}{S.~Zhu},
  \bibinfo{author}{S.~S. Ghugre}, \bibinfo{author}{R.~Palit},
  \bibinfo{journal}{Phys. Rev. Lett.} \bibinfo{volume}{110}
  (\bibinfo{year}{2013}) \bibinfo{pages}{172504}.
\bibitem[{Frauendorf(2001)}]{Frauendorf2001Rev.Mod.Phys.463514}
\bibinfo{author}{S.~Frauendorf}, \bibinfo{journal}{Rev. Mod. Phys.}
  \bibinfo{volume}{73} (\bibinfo{year}{2001}) \bibinfo{pages}{463--514}.
\bibitem[{Meng and
  Zhang(2010)}]{Meng2010JournalofPhysicsGNuclearandParticlePhysics064025}
\bibinfo{author}{J.~Meng}, \bibinfo{author}{S.~Q. Zhang}, \bibinfo{journal}{J.
  Phys. G: Nucl. Part. Phys.} \bibinfo{volume}{37} (\bibinfo{year}{2010})
  \bibinfo{pages}{064025}.
\bibitem[{Meng and Zhao(2016)}]{Meng2016PhysicaScripta053008}
\bibinfo{author}{J.~Meng}, \bibinfo{author}{P.~W. Zhao},
  \bibinfo{journal}{Phys. Scr.} \bibinfo{volume}{91} (\bibinfo{year}{2016})
  \bibinfo{pages}{053008}.
\bibitem[{Xiong and Wang(2019)}]{Xiong2019AtomicDataandNuclearDataTables193225}
\bibinfo{author}{B.~W. Xiong}, \bibinfo{author}{Y.~Y. Wang},
  \bibinfo{journal}{At. Data Nucl. Data Tables} \bibinfo{volume}{125}
  (\bibinfo{year}{2019}) \bibinfo{pages}{193--225}.
\bibitem[{Koike et~al.(2004)Koike, Starosta, and
  Hamamoto}]{Koike2004Phys.Rev.Lett.172502}
\bibinfo{author}{T.~Koike}, \bibinfo{author}{K.~Starosta},
  \bibinfo{author}{I.~Hamamoto}, \bibinfo{journal}{Phys. Rev. Lett.}
  \bibinfo{volume}{93} (\bibinfo{year}{2004}) \bibinfo{pages}{172502}.
\bibitem[{Peng et~al.(2003)Peng, Meng, and Zhang}]{Peng2003Phys.Rev.C044324}
\bibinfo{author}{J.~Peng}, \bibinfo{author}{J.~Meng}, \bibinfo{author}{S.~Q.
  Zhang}, \bibinfo{journal}{Phys. Rev. C} \bibinfo{volume}{68}
  (\bibinfo{year}{2003}) \bibinfo{pages}{044324}.
\bibitem[{Zhang et~al.(2007)Zhang, Qi, Wang, and
  Meng}]{Zhang2007Phys.Rev.C044307}
\bibinfo{author}{S.~Q. Zhang}, \bibinfo{author}{B.~Qi}, \bibinfo{author}{S.~Y.
  Wang}, \bibinfo{author}{J.~Meng}, \bibinfo{journal}{Phys. Rev. C}
  \bibinfo{volume}{75} (\bibinfo{year}{2007}) \bibinfo{pages}{044307}.
\bibitem[{Qi et~al.(2009)Qi, Zhang, Meng, Wang, and
  Frauendorf}]{Qi2009PhysicsLettersB175180}
\bibinfo{author}{B.~Qi}, \bibinfo{author}{S.~Q. Zhang},
  \bibinfo{author}{J.~Meng}, \bibinfo{author}{S.~Y. Wang},
  \bibinfo{author}{S.~Frauendorf}, \bibinfo{journal}{Phys. Lett. B}
  \bibinfo{volume}{675} (\bibinfo{year}{2009}) \bibinfo{pages}{175--180}.
\bibitem[{Chen et~al.(2018)Chen, Lv, Petrache, and
  Meng}]{Chen2018PhysicsLettersB744749}
\bibinfo{author}{Q.~B. Chen}, \bibinfo{author}{B.~F. Lv},
  \bibinfo{author}{C.~M. Petrache}, \bibinfo{author}{J.~Meng},
  \bibinfo{journal}{Phys. Lett. B} \bibinfo{volume}{782} (\bibinfo{year}{2018})
  \bibinfo{pages}{744--749}.
\bibitem[{Raduta et~al.(2016)Raduta, Raduta, and
  Petrache}]{Raduta2016JournalofPhysicsGNuclearandParticlePhysics095107}
\bibinfo{author}{A.~A. Raduta}, \bibinfo{author}{A.~H. Raduta},
  \bibinfo{author}{C.~M. Petrache}, \bibinfo{journal}{J. Phys. G: Nucl. Part.
  Phys.} \bibinfo{volume}{43} (\bibinfo{year}{2016}) \bibinfo{pages}{095107}.
\bibitem[{Tonev et~al.(2006)Tonev, de~Angelis, Petkov, Dewald, Brant,
  Frauendorf, Balabanski, Pejovic, Bazzacco, Bednarczyk, Camera, Fitzler,
  Gadea, Lenzi, Lunardi, Marginean, M\"oller, Napoli, Paleni, Petrache, Prete,
  Zell, Zhang, Zhang, Zhong, and Curien}]{Tonev2006Phys.Rev.Lett.052501}
\bibinfo{author}{D.~Tonev}, \bibinfo{author}{G.~de~Angelis},
  \bibinfo{author}{P.~Petkov}, \bibinfo{author}{A.~Dewald},
  \bibinfo{author}{S.~Brant}, \bibinfo{author}{S.~Frauendorf},
  \bibinfo{author}{D.~L. Balabanski}, \bibinfo{author}{P.~Pejovic},
  \bibinfo{author}{D.~Bazzacco}, \bibinfo{author}{P.~Bednarczyk},
  \bibinfo{author}{F.~Camera}, \bibinfo{author}{A.~Fitzler},
  \bibinfo{author}{A.~Gadea}, \bibinfo{author}{S.~Lenzi},
  \bibinfo{author}{S.~Lunardi}, \bibinfo{author}{N.~Marginean},
  \bibinfo{author}{O.~M\"oller}, \bibinfo{author}{D.~R. Napoli},
  \bibinfo{author}{A.~Paleni}, \bibinfo{author}{C.~M. Petrache},
  \bibinfo{author}{G.~Prete}, \bibinfo{author}{K.~O. Zell},
  \bibinfo{author}{Y.~H. Zhang}, \bibinfo{author}{J.~Y. Zhang},
  \bibinfo{author}{Q.~Zhong}, \bibinfo{author}{D.~Curien},
  \bibinfo{journal}{Phys. Rev. Lett.} \bibinfo{volume}{96}
  (\bibinfo{year}{2006}) \bibinfo{pages}{052501}.
\bibitem[{Dimitrov et~al.(2000)Dimitrov, Frauendorf, and
  D\"onau}]{Dimitrov2000Phys.Rev.Lett.57325735}
\bibinfo{author}{V.~I. Dimitrov}, \bibinfo{author}{S.~Frauendorf},
  \bibinfo{author}{F.~D\"onau}, \bibinfo{journal}{Phys. Rev. Lett.}
  \bibinfo{volume}{84} (\bibinfo{year}{2000}) \bibinfo{pages}{5732--5735}.
\bibitem[{Chen et~al.(2013)Chen, Zhang, Zhao, Jolos, and
  Meng}]{Chen2013Phys.Rev.C024314}
\bibinfo{author}{Q.~B. Chen}, \bibinfo{author}{S.~Q. Zhang},
  \bibinfo{author}{P.~W. Zhao}, \bibinfo{author}{R.~V. Jolos},
  \bibinfo{author}{J.~Meng}, \bibinfo{journal}{Phys. Rev. C}
  \bibinfo{volume}{87} (\bibinfo{year}{2013}) \bibinfo{pages}{024314}.
\bibitem[{Bhat et~al.(2012)Bhat, Sheikh, and
  Palit}]{Bhat2012PhysicsLettersB250254}
\bibinfo{author}{G.~H. Bhat}, \bibinfo{author}{J.~A. Sheikh},
  \bibinfo{author}{R.~Palit}, \bibinfo{journal}{Phys. Lett. B}
  \bibinfo{volume}{707} (\bibinfo{year}{2012}) \bibinfo{pages}{250--254}.
\bibitem[{Chen et~al.(2017)Chen, Chen, Luo, Meng, and
  Zhang}]{Chen2017Phys.Rev.C051303}
\bibinfo{author}{F.~Q. Chen}, \bibinfo{author}{Q.~B. Chen},
  \bibinfo{author}{Y.~A. Luo}, \bibinfo{author}{J.~Meng},
  \bibinfo{author}{S.~Q. Zhang}, \bibinfo{journal}{Phys. Rev. C}
  \bibinfo{volume}{96} (\bibinfo{year}{2017}) \bibinfo{pages}{051303}.
\bibitem[{Chen et~al.(2018)Chen, Meng, and
  Zhang}]{Chen2018PhysicsLettersB211216}
\bibinfo{author}{F.~Q. Chen}, \bibinfo{author}{J.~Meng}, \bibinfo{author}{S.~Q.
  Zhang}, \bibinfo{journal}{Phys. Lett. B} \bibinfo{volume}{785}
  (\bibinfo{year}{2018}) \bibinfo{pages}{211--216}.
\bibitem[{Wang et~al.(2019)Wang, Chen, Zhao, Zhang, and
  Meng}]{Wang2019Phys.Rev.C054303}
\bibinfo{author}{Y.~K. Wang}, \bibinfo{author}{F.~Q. Chen},
  \bibinfo{author}{P.~W. Zhao}, \bibinfo{author}{S.~Q. Zhang},
  \bibinfo{author}{J.~Meng}, \bibinfo{journal}{Phys. Rev. C}
  \bibinfo{volume}{99} (\bibinfo{year}{2019}) \bibinfo{pages}{054303}.
\bibitem[{Olbratowski et~al.(2004)Olbratowski, Dobaczewski, Dudek, and
  P\l{}\'ociennik}]{Olbratowski2004Phys.Rev.Lett.052501}
\bibinfo{author}{P.~Olbratowski}, \bibinfo{author}{J.~Dobaczewski},
  \bibinfo{author}{J.~Dudek}, \bibinfo{author}{W.~P\l{}\'ociennik},
  \bibinfo{journal}{Phys. Rev. Lett.} \bibinfo{volume}{93}
  (\bibinfo{year}{2004}) \bibinfo{pages}{052501}.
\bibitem[{Zhao(2017)}]{Zhao2017PhysicsLettersB15}
\bibinfo{author}{P.~W. Zhao}, \bibinfo{journal}{Phys. Lett. B}
  \bibinfo{volume}{773} (\bibinfo{year}{2017}) \bibinfo{pages}{1--5}.
\bibitem[{Zhao et~al.(2019)Zhao, Wang, and Chen}]{Zhao2019Phys.Rev.C054319}
\bibinfo{author}{P.~W. Zhao}, \bibinfo{author}{Y.~K. Wang},
  \bibinfo{author}{Q.~B. Chen}, \bibinfo{journal}{Phys. Rev. C}
  \bibinfo{volume}{99} (\bibinfo{year}{2019}) \bibinfo{pages}{054319}.
\bibitem[{Peng and Chen(2020)}]{Peng2020PhysicsLettersB135795}
\bibinfo{author}{J.~Peng}, \bibinfo{author}{Q.~B. Chen},
  \bibinfo{journal}{Phys. Lett. B} \bibinfo{volume}{810} (\bibinfo{year}{2020})
  \bibinfo{pages}{135795}.
\bibitem[{Wang and Meng(2023)}]{Wang2023PhysicsLettersB137923}
\bibinfo{author}{Y.~P. Wang}, \bibinfo{author}{J.~Meng},
  \bibinfo{journal}{Phys. Lett. B} \bibinfo{volume}{841} (\bibinfo{year}{2023})
  \bibinfo{pages}{137923}.
\bibitem[{Meng(2016)}]{Meng2016}
\bibinfo{editor}{J.~Meng} (Ed.), \bibinfo{title}{Relativistic Density
  Functional for Nuclear Structure}, volume~\bibinfo{volume}{10} of
  \textit{\bibinfo{series}{International Review of Nuclear Physics}},
  \bibinfo{publisher}{World Scientific, Singapore}, \bibinfo{year}{2016}.
\bibitem[{Ren et~al.(2022)Ren, Zhao, and Meng}]{Ren2022Phys.Rev.CL011301}
\bibinfo{author}{Z.~X. Ren}, \bibinfo{author}{P.~W. Zhao},
  \bibinfo{author}{J.~Meng}, \bibinfo{journal}{Phys. Rev. C}
  \bibinfo{volume}{105} (\bibinfo{year}{2022}) \bibinfo{pages}{L011301}.
\bibitem[{Zhao et~al.(2016)Zhao, Ring, and Meng}]{Zhao2016Phys.Rev.C041301}
\bibinfo{author}{P.~W. Zhao}, \bibinfo{author}{P.~Ring},
  \bibinfo{author}{J.~Meng}, \bibinfo{journal}{Phys. Rev. C}
  \bibinfo{volume}{94} (\bibinfo{year}{2016}) \bibinfo{pages}{041301}.
\bibitem[{Wang et~al.(2023)Wang, Zhao, and
  Meng}]{Wang2023arXivpreprintarXiv2304.12009}
\bibinfo{author}{Y.~K. Wang}, \bibinfo{author}{P.~W. Zhao},
  \bibinfo{author}{J.~Meng}, \bibinfo{journal}{arXiv:2304.12009}
  (\bibinfo{year}{2023}).
\bibitem[{Starosta et~al.(2001)Starosta, Koike, Chiara, Fossan, LaFosse, Hecht,
  Beausang, Caprio, Cooper, Kr\"ucken, Novak, Zamfir, Zyromski, Hartley,
  Balabanski, Zhang, Frauendorf, and
  Dimitrov}]{Starosta2001Phys.Rev.Lett.971974}
\bibinfo{author}{K.~Starosta}, \bibinfo{author}{T.~Koike},
  \bibinfo{author}{C.~J. Chiara}, \bibinfo{author}{D.~B. Fossan},
  \bibinfo{author}{D.~R. LaFosse}, \bibinfo{author}{A.~A. Hecht},
  \bibinfo{author}{C.~W. Beausang}, \bibinfo{author}{M.~A. Caprio},
  \bibinfo{author}{J.~R. Cooper}, \bibinfo{author}{R.~Kr\"ucken},
  \bibinfo{author}{J.~R. Novak}, \bibinfo{author}{N.~V. Zamfir},
  \bibinfo{author}{K.~E. Zyromski}, \bibinfo{author}{D.~J. Hartley},
  \bibinfo{author}{D.~L. Balabanski}, \bibinfo{author}{J.~Y. Zhang},
  \bibinfo{author}{S.~Frauendorf}, \bibinfo{author}{V.~I. Dimitrov},
  \bibinfo{journal}{Phys. Rev. Lett.} \bibinfo{volume}{86}
  (\bibinfo{year}{2001}) \bibinfo{pages}{971--974}.
\bibitem[{Wang et~al.(2022)Wang, Zhao, and Meng}]{Wang2022Phys.Rev.C054311}
\bibinfo{author}{Y.~K. Wang}, \bibinfo{author}{P.~W. Zhao},
  \bibinfo{author}{J.~Meng}, \bibinfo{journal}{Phys. Rev. C}
  \bibinfo{volume}{105} (\bibinfo{year}{2022}) \bibinfo{pages}{054311}.
\bibitem[{Ring and Schuck(2004)}]{Ring2004}
\bibinfo{author}{P.~Ring}, \bibinfo{author}{P.~Schuck}, \bibinfo{title}{The
  nuclear many-body problem}, \bibinfo{publisher}{Springer Science \& Business
  Media, New York}, \bibinfo{year}{2004}.
\bibitem[{Hu et~al.(2014)Hu, Gao, and Chen}]{Hu2014PhysicsLettersB162166}
\bibinfo{author}{Q.~L. Hu}, \bibinfo{author}{Z.~C. Gao},
  \bibinfo{author}{Y.~Chen}, \bibinfo{journal}{Phys. Lett. B}
  \bibinfo{volume}{734} (\bibinfo{year}{2014}) \bibinfo{pages}{162--166}.
\bibitem[{Carlsson and Rotureau(2021)}]{Carlsson2021Phys.Rev.Lett.172501}
\bibinfo{author}{B.~G. Carlsson}, \bibinfo{author}{J.~Rotureau},
  \bibinfo{journal}{Phys. Rev. Lett.} \bibinfo{volume}{126}
  (\bibinfo{year}{2021}) \bibinfo{pages}{172501}.
\bibitem[{Zhao et~al.(2010)Zhao, Li, Yao, and Meng}]{Zhao2010Phys.Rev.C054319}
\bibinfo{author}{P.~W. Zhao}, \bibinfo{author}{Z.~P. Li},
  \bibinfo{author}{J.~M. Yao}, \bibinfo{author}{J.~Meng},
  \bibinfo{journal}{Phys. Rev. C} \bibinfo{volume}{82} (\bibinfo{year}{2010})
  \bibinfo{pages}{054319}.
\bibitem[{Tian et~al.(2009)Tian, Ma, and Ring}]{Tian2009PhysicsLettersB4450}
\bibinfo{author}{Y.~Tian}, \bibinfo{author}{Z.~Y. Ma},
  \bibinfo{author}{P.~Ring}, \bibinfo{journal}{Physics Letters B}
  \bibinfo{volume}{676} (\bibinfo{year}{2009}) \bibinfo{pages}{44--50}.
\bibitem[{Nik{\v{s}}i{\'c} et~al.(2014)Nik{\v{s}}i{\'c}, Paar, Vretenar, and
  Ring}]{Niksic2014ComputerPhysicsCommunications18081821}
\bibinfo{author}{T.~Nik{\v{s}}i{\'c}}, \bibinfo{author}{N.~Paar},
  \bibinfo{author}{D.~Vretenar}, \bibinfo{author}{P.~Ring},
  \bibinfo{journal}{Comput. Phys. Commun.} \bibinfo{volume}{185}
  (\bibinfo{year}{2014}) \bibinfo{pages}{1808--1821}.
\bibitem[{Wang et~al.(2009)Wang, Wu, Zhu, Li, Hao, Zheng, He, Wang, Li, Liu,
  Pan, Li, and Ding}]{LieLin2009ChinesePhysicsC173}
\bibinfo{author}{L.~L. Wang}, \bibinfo{author}{X.~G. Wu},
  \bibinfo{author}{L.~H. Zhu}, \bibinfo{author}{G.~S. Li},
  \bibinfo{author}{X.~Hao}, \bibinfo{author}{Y.~Zheng}, \bibinfo{author}{C.~Y.
  He}, \bibinfo{author}{L.~Wang}, \bibinfo{author}{X.~Q. Li},
  \bibinfo{author}{Y.~Liu}, \bibinfo{author}{B.~Pan}, \bibinfo{author}{Z.~Y.
  Li}, \bibinfo{author}{H.~B. Ding}, \bibinfo{journal}{Chin. Phys. C}
  \bibinfo{volume}{33} (\bibinfo{year}{2009}) \bibinfo{pages}{173}.
\bibitem[{Qi et~al.(2009)Qi, Zhang, Wang, Yao, and
  Meng}]{Qi2009Phys.Rev.C041302}
\bibinfo{author}{B.~Qi}, \bibinfo{author}{S.~Q. Zhang}, \bibinfo{author}{S.~Y.
  Wang}, \bibinfo{author}{J.~M. Yao}, \bibinfo{author}{J.~Meng},
  \bibinfo{journal}{Phys. Rev. C} \bibinfo{volume}{79} (\bibinfo{year}{2009})
  \bibinfo{pages}{041302}.

\end{thebibliography}
\end{document}